\documentclass{moriond_modified}

\usepackage{wrapfig}
\usepackage{amssymb}
\usepackage{hyperref}

\def\Journal#1#2#3#4{{#1} {\bf #2}, #3 (#4)}

\def\PRL{\em Phys. Rev. Lett.}
\def\PRD{{\em Phys. Rev.} D}

\def\AA{{\em Astron. Astrophys.}}
\def\ApJ{{\em Astrophys. J.}}
\def\ApJS{{\em Astrophys. J. Supp.}}

\def\be{\begin{equation}}
\def\ee{\end{equation}}
\def\bea{\begin{eqnarray}}
\def\eea{\end{eqnarray}}

\def\bk{BICEP/\textit{Keck}}
\newcommand{\Photo}{}

\begin{document}
\title{Constraining Inflation with the BICEP/\textit{Keck} CMB
  Polarization Experiments}

\author[a]{The \bk\ Collaboration: P.~A.~R.~Ade}
\author[b]{Z.~Ahmed}
\author[c]{M.~Amiri}
\author[d]{D.~Barkats}
\author[e,h]{R.~Basu~Thakur}
\author[f]{C.~A.~Bischoff}
\author[b,g]{D.~Beck}
\author[e,h]{J.~J.~Bock}
\author[d]{H.~Boenish}
\author[i]{V.~Buza}
\author[j,*]{J.~R.~Cheshire~IV}
\author[d]{J.~Connors}
\author[d]{J.~Cornelison}
\author[k]{M.~Crumrine}
\author[e]{A.~Cukierman}
\author[l]{E.~V.~Denison}
\author[d]{M.~Dierickx}
\author[m]{L.~Duband}
\author[d]{M.~Eiben}
\author[d]{B.~Elwood}
\author[c,e]{S.~Fatigoni}
\author[n,o]{J.~P.~Filippini}
\author[e]{M.~Gao}
\author[f]{C.~Giannakopoulos}
\author[g]{N.~Goeckner-Wald}
\author[b]{D.~C.~Goldfinger} % becomes "p" from "d"
\author[g]{J.~Grayson}
\author[d]{P.~Grimes}
\author[k]{G.~Hall}
\author[g]{G.~Halal}
\author[c]{M.~Halpern}
\author[f]{E.~Hand}
\author[d]{S.~Harrison}
\author[b]{S.~Henderson}
\author[l]{J.~Hubmayr}
\author[e]{H.~Hui}
\author[g,b,l]{K.~D.~Irwin}
\author[g,e]{J.~Kang}
\author[b]{K.~S.~Karkare}
\author[e]{S.~Kefeli}
\author[d,p]{J.~M.~Kovac}
\author[g,b]{C.~L.~Kuo}
\author[e]{K.~Lau}
\author[n]{A.~Lennox}
\author[g]{T.~Liu}
\author[h]{K.~G.~Megerian}
\author[e]{L.~Minutolo}
\author[e]{L.~Moncelsi}
\author[g]{Y.~Nakato}
\author[q]{T.~Namikawa}
\author[h]{H.~T.~Nguyen}
\author[e,h]{R.~O'Brient}
\author[f]{S.~Palladino}
\author[d]{M.~Petroff}
\author[d]{A.~Polish}
\author[k]{N.~Precup}
\author[m]{T.~Prouve}
\author[k,j]{C.~Pryke}
\author[d,r]{B.~Racine}
\author[l]{C.~D.~Reintsema}
\author[e]{T.~Romand}
\author[g]{M.~Salatino}
\author[e]{A.~Schillaci}
\author[d,s]{B.~L.~Schmitt}
\author[j]{B.~Singari}
\author[e]{A.~Soliman}
\author[d,p]{T.~St.~Germaine}
\author[e]{A.~Steiger}
\author[e]{B.~Steinbach}
\author[a]{R.~V.~Sudiwala}
\author[g,b]{K.~L.~Thompson}
\author[a]{C.~Tucker}
\author[h]{A.~D.~Turner}
\author[d]{C.~Verg\`{e}s}
\author[t,j]{A.~G.~Vieregg}
\author[e]{A.~Wandui}
\author[h]{A.~C.~Weber}
\author[k]{J.~Willmert}
\author[b]{W.~L.~K.~Wu}
\author[g]{H.~Yang}
\author[g,b]{K.~W.~Yoon}
\author[g,b]{E.~Young}
\author[g]{C.~Yu}
\author[d]{L.~Zeng}
\author[g,e]{C.~Zhang}
\author[e]{S.~Zhang}

\affil[a]{School of Physics and Astronomy, Cardiff University, Cardiff, CF24 3AA, United Kingdom}
\affil[b]{Kavli Institute for Particle Astrophysics and Cosmology, SLAC National Accelerator Laboratory, 2575 Sand Hill Rd, Menlo Park, CA 94025, USA}
\affil[c]{Department of Physics and Astronomy, University of British Columbia, Vancouver, British Columbia, V6T 1Z1, Canada}
\affil[d]{Center for Astrophysics, Harvard \& Smithsonian, Cambridge, MA 02138, USA}
\affil[e]{Department of Physics, California Institute of Technology, Pasadena, CA 91125, USA}
\affil[f]{Department of Physics, University of Cincinnati, Cincinnati, OH 45221, USA}
\affil[g]{Department of Physics, Stanford University, Stanford, CA 94305, USA}
\affil[h]{Jet Propulsion Laboratory, Pasadena, CA 91109, USA}
\affil[i]{Kavli Institute for Cosmological Physics, University of Chicago, Chicago, IL 60637, USA}
\affil[j]{Minnesota Institute for Astrophysics, University of Minnesota, Minneapolis, MN 55455, USA}
\affil[k]{School of Physics and Astronomy, University of Minnesota, Minneapolis, MN 55455, USA}
\affil[l]{National Institute of Standards and Technology, Boulder, CO 80305, USA}
\affil[m]{Service des Basses Temp\'{e}ratures, Commissariat \`{a} l'Energie Atomique, 38054 Grenoble, France}
\affil[n]{Department of Physics, University of Illinois at Urbana-Champaign, Urbana, IL 61801, USA}
\affil[o]{Department of Astronomy, University of Illinois at Urbana-Champaign, Urbana, IL 61801, USA}

\affil[p]{Department of Physics, Harvard University, Cambridge, MA 02138, USA}
\affil[q]{Kavli Institute for the Physics and Mathematics of the Universe (WPI), UTIAS, The~University~of~Tokyo, Kashiwa, Chiba 277-8583, Japan}
\affil[r]{Aix-Marseille  Universit\'{e},  CNRS/IN2P3,  CPPM,  13288 Marseille,  France}
\affil[s]{Department of Physics and Astronomy, University of Pennsylvania, Philadelphia, PA 19104, USA}
\affil[t]{Department of Physics, Enrico Fermi Institute, University of Chicago, Chicago, IL 60637, USA}

\authorinfo{*Corresponding author: J.R.~Cheshire~IV, \href{mailto:cheshire@umn.edu}{cheshire@umn.edu}}

\maketitle\abstracts{
The BICEP/\textit{Keck} (BK) series of cosmic microwave background
(CMB) polarization experiments has, over the past decade and a half,
produced a series of field-leading constraints on cosmic
inflation via measurements of the ``B-mode'' polarization of the
CMB. Primordial B modes are directly tied to the amplitude of
primordial gravitational waves (PGW), thier strength parameterized by
the tensor-to-scalar ratio, $r$, and thus the energy scale of
inflation. Having set the most sensitive constraints to-date on
$r$, $\sigma(r)=0.009$ ($r_{0.05}<0.036,
95\%$ C.L.) using data through the 2018 observing season (``BK18''),
the BICEP/\textit{Keck} program has continued to improve its dataset
in the years since. We give a brief overview of the BK program and the
``BK18'' result before discussing the program's
ongoing efforts, including the deployment and performance of the
\textit{Keck Array}'s successor instrument, BICEP Array, improvements
to data processing and internal consistency testing, new techniques
such as delensing, and how those will ultimately serve to allow BK
reach $\sigma(r) \lesssim  0.003$ using data through the 2027
observing season.}

\section{Introduction}
\label{sec:intro}

Our Universe is well-described by a hot Big Bang model, with the bulk
of its
energy density made up by a cold dark matter (CDM) component and a
dominant cosmological constant component
$\Lambda$~\cite{p2018cosmo}. This ``$\Lambda$CDM'' model has seen
extensive
success in matching observations of both the low- and high-redshift
universes, but, on its own, possesses a few noteworthy
deficiencies. It does not provide a mechanism for the generation of
the primordial perturbations which seeded further structure growth,
nor does it explain the thermalization of the CMB on acausal scales
(the ``horizon problem'') or the apparent fine-tuning of the
cosmological spatial flatness parameter (the ``flatness
problem''). These deficiencies can be addressed through the inclusion of a period of rapid
accelerating expansion, ``inflation'', at very early times. This
generic paradigm has seen a wealth of indirect evidence,
with the lowest-order predicitions of Gaussianity, adiabaticity, and
near scale-invariance of the primoridal perturbation spectrum all
confirmed to exquisite precision~\cite{p2018inflation}. However,
inflation also predicts a background of PGW~\cite{starobinskypgw},
as-yet undetected. If these PGW could be measured, this would serve as
direct evidence --- a ``smoking gun'' --- for the
inflationary scenario.

Such a background of PGW would induce a particular odd-parity
``B-mode'' pattern in the polarization of the
CMB~\cite{bmode1,bmode2}, which
may be detectable. $\Lambda$CDM (scalar) fluctuations are
only capable of generating even-parity ``E modes'', thus a detection
of primordial B modes would be tantamount to a detection of PGW, and a
direct probe of the inflationary potential. Temperature-based
probes of inflation have become limited by cosmic variance in the past
decade~\cite{p2018inflation}, thus B-mode polarization is currently
the most promising avenue by which to search for direct evidence of
inflation.

Practical challenges to measuring primordial B-mode polarization are
manyfold. Due to the weak nature of any potential primordial B-mode
signal, temperature-to-polarization leakage introduced by the
instrument and E-to-B leakage introduced by the experiment design and analysis may be major concerns, and weak
instrumental systematics may become critically important. There are
also significant astrophysical challenges, in the form of foreground
emission (primarily from galactic dust, but potentially also due to
mechanisms such as galactic synchrotron radiation) and gravitational
lensing of the CMB by large-scale structure along the line-of-sight,
which mixes E and B and must be carefully accounted for (see
subsection~\ref{subsec:delensing} and references therein).

Herein, we proceed by briefly reviewing the experimental strategy and
history of the \bk\ program to-date in section~\ref{sec:experiment},
briefly noting the current cosmological constraints in
section~\ref{sec:bk18}, and then in section~\ref{sec:newdata} discussing the progress of the \bk\
program since the previous publication. Finally, we review the future
directions of the project in section~\ref{sec:future}.

\section{The BICEP/\textit{Keck}~Experiments}
\label{sec:experiment}

The \bk\ telescopes search for B-mode polarization by employing
small-aperture refracting polarimeters at the geographic South Pole to
observe a small patch of sky continuously throughout the Polar
winter. Utilizing small-aperture telescopes enables rigorous control
of instrumental systematic effects, ease of maintenance and
upgradability, and facilitates boresight rotation. BK receivers
utilize superconducting transition edge sensor (TES) bolometer arrays~\cite{bkspidertes} and
all-cold refractive optics, and have significant internal and external
baffling/shielding to mitigate off-axis coupling of the receiver to
terrestrial sources of radiation.

%BK
%telescopes have the ability to rotate around the line-of-sight,
%thereby modulating the absolute polarization angle of the detectors on
%the sky, which can suppress some forms of polarized systematics and
%allows for the robust reconstruction of Stokes $Q$/$U$ without any
%polarizing optical elements.
%
%The telescopes themselves utilize transition edge sensor (TES)
%bolometers coupled to slot-dipole phased antenna array
%networks~\cite{bkspidertes}. Each focal plane ``pixel'' has two
%interwoven, co-located, antenna arrays measuring orthogonal
%polarization modes, the difference of their signals therefore being a
%measure of linear polarization. These detectors are deployed in large
%planar arrays, referred to as detector ``tiles'', which for BICEP3 and
%BICEP Array onwards are integrated into modular focal plane units (``modules'')
%containing additional first-stage amplification and readout, and
%enabling easier and more rapid swapping of detector elements. These
%focal planes are housed in cryostats containing cold (4K) lens and
%filtering elements, with the focal plane itself cooled to
%$<300$ mK. Significant internal and
%external baffling and external shielding serves to control stray light and
%off-axis coupling of the telescope receivers to terrestrial sources of radiation.

\bk\ telescopes scan a constant throw in azimuth, periodically
(once every $\sim2$ hours) updating the Right Ascension to center on
the observing field. Allowing the observing field to drift with the
rotation of the Earth over the
set of scans allows for robust removal of azimuth-fixed
signals by removing the common signal in azimuth across all
scans. After a fixed period of scans and integrated calibration
measurements, the helium sorption refrigerator
cooling the focal plane is cycled and any routine maintenance is
performed, before beginning a new set of scans. This process is
repeated constantly throughout the Austral winter and ultimately over
multiple years (observing seasons).

\begin{wrapfigure}{r}{0.5\textwidth}
  \centering
  \includegraphics[width=0.5\textwidth]{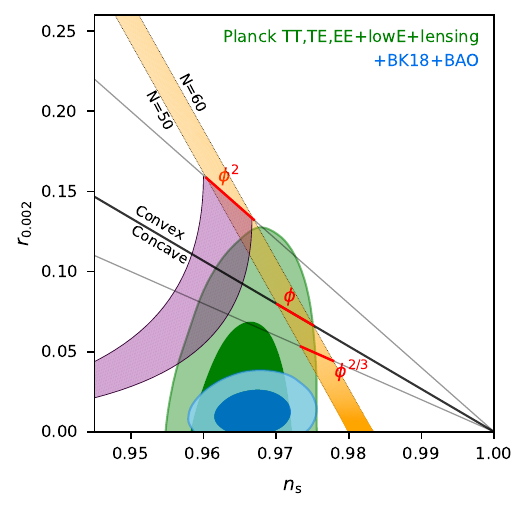}
  \caption{Constraints in the $r-n_s$ plane with the inclusion of \bk\
  and BAO data. Note that BAO data primarily results in the shrinking
  in the $n_s$-direction; the improvement in $r$ is driven almost
  entirely by \bk. Adapted from Figure 5 of BK18~\protect\cite{BK18}.}
  \label{fig:r_vs_ns_bk18}
\end{wrapfigure}

BICEP2 observed from 2010--2012 at 150 GHz. \textit{Keck Array} was an
array of five BICEP2-class receivers operating from 2012--2019,
later observing at multiple frequencies
from 95 to 270 GHz to discriminate foregrounds (dust, for example,
being brighter at higher frequencies). BICEP3 began science observations in 2016 at 95
GHz, featuring an instantaneous sensitivity comparable to the entire
\textit{Keck Array}. BICEP Array,
an array of BICEP3-class receivers succeeding \textit{Keck}, began
operations along with its first receiver in 2020.

\section{BK18 Cosmological Constraints}
\label{sec:bk18}

%\begin{wrapfigure}{r}{0.5\textwidth}
%  \centering
%  \includegraphics[width=0.5\textwidth]{r_vs_ns_bk18}
%  \caption{Constraints in the $r-n_s$ plane with the inclusion of \bk\
%  and BAO data. Note that BAO data primarily results in the shrinking
%  in the $n_s$-direction; the improvement in $r$ is driven almost
%  entirely by \bk. Adapted from Figure 5 of BK18~\protect\cite{BK18}.}
%  \label{fig:r_vs_ns_bk18}
%\end{wrapfigure}

\bk's latest published ``BK18'' constraints~\cite{BK18} are based on data up-to
and including the 2018 observing season, in addition to external data
from \textit{Planck} and WMAP, and are the first published
constraints to include data taken by BICEP3. BK18 set the limit
$r_{0.05}<0.036$ ($95\%$ C. L.), with a sensitivity
$\sigma(r)=0.009$. The constraint on the $r-n_s$ plane is shown in
figure~\ref{fig:r_vs_ns_bk18}. These remain the most sensitive published
constraints on primordial gravitational waves from CMB B-mode
polarization. BK18's sensitivity is dominated by only three seasons of BICEP3 data,
and lensing sample variance is now the dominant contribution to BK18's
$\sigma(r)$, rather than the uncertainty on foreground emission as in
prior releases. These
last two facts serve to underscore the importance of BK's ongoing
efforts, described in the subsequent sections.

BK18 employs BK's historical standard analysis procedure of
constructing an equidistant cylindrical projection map, estimating the
angular power spectra based on the mean in annuli of the 2-D Fourier
transforms of those maps, and evaluating the joint likelihood of all resulting auto- and
cross-spectra against a multicomponent model consisting of
lensed-$\Lambda$CDM, foregrounds, and $r$. Extensive suites of
simulations were constructed to both estimate fluctuation and significance
levels and against which to compare various data splits to probe for
systematics.

\section{Data Collected Since 2018}
\label{sec:newdata}

Since 2018, BICEP3 and its
companion instrument --- \textit{Keck Array}, succeeded by
BICEP~Array --- have continued to observe the CMB during each year's
Austral winter season. This represents a significant increase in
survey weight at 95 GHz, and the extension of the \bk~program's
frequency coverage to both higher and lower observing frequencies. The
next \bk\ map set is currently planned to be ``BK23'', containing data
up to and including 2023. We briefly describe the progress and show
preliminary BK23 coadded maps for a noteworthy subset of frequency bands.

\subsection{BICEP3}
\label{subsec:b3}

BICEP3 was deployed during the 2014--2015 Austral summer season,
beginning science observations during the 2016 Austral winter
season. BICEP3 was improved significantly
between the 2016 and 2017 seasons with the installation of
better-performing detector tiles and the replacement of some optical 
filtering elements. BICEP3 and the three-year dataset are described in
detail in {\em \bk~Collaboration et al.}~(2022)~\cite{b3inst}. During
the 2022-2023 Austral summer, the computer
systems for BICEP3, inherited from BICEP2 and nearing 15 years of age,
were replaced, and the same season an experimental ultra-thin high
modulus polyethylene (HMPE)
vacuum window~\cite{mirandawindow} was installed to improve
loading and overall sensitivity. Otherwise, BICEP3 has
operated in a configuration identical to that described in the BICEP3
instrument paper~\cite{b3inst}.

\begin{figure}[t]
  \centering
  \includegraphics[width=1\textwidth]{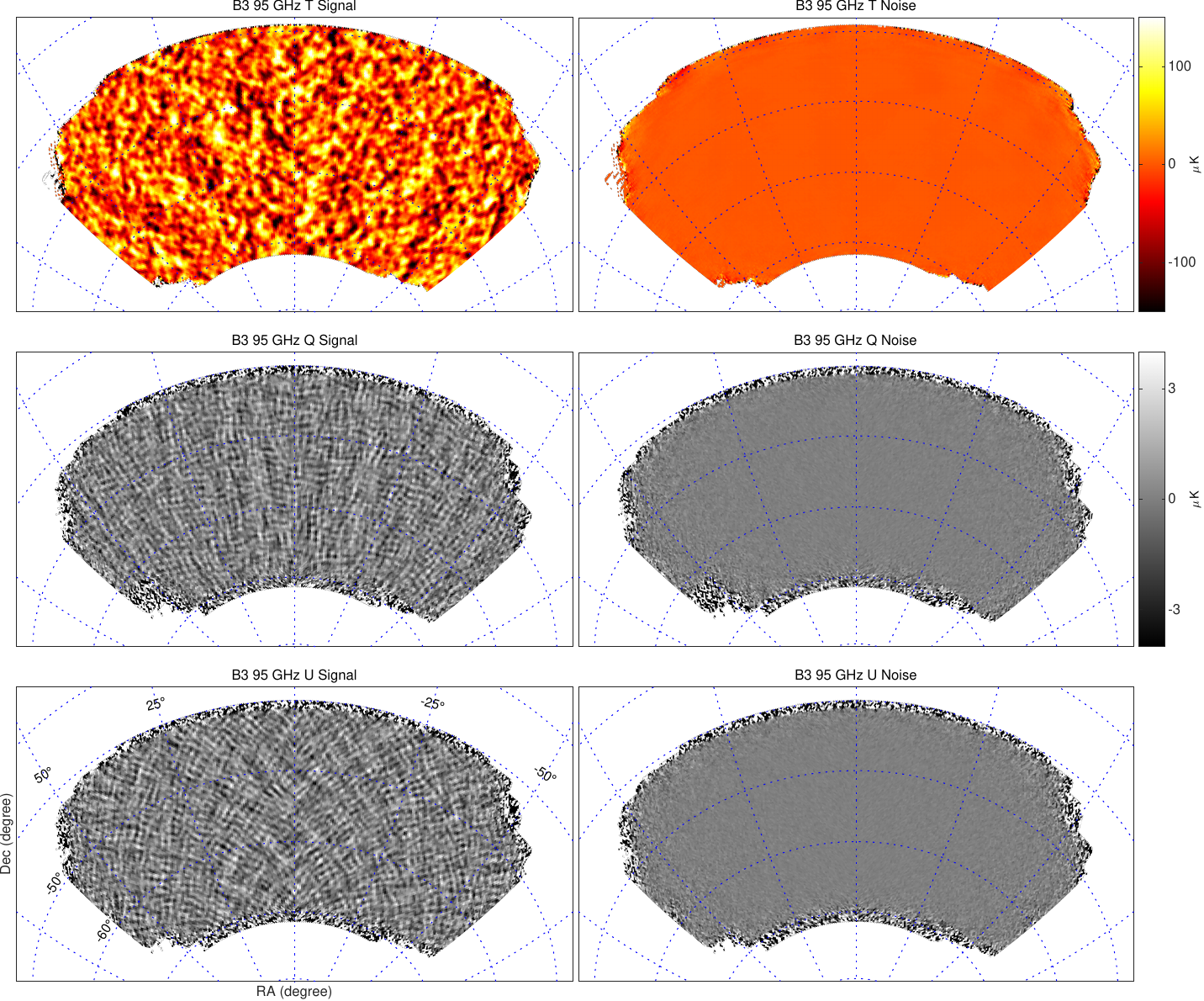}
  \caption{Preliminary BICEP3 95 GHz 8-year (``BK23 L95'') temperature and
    polarization maps.}
\label{fig:b3_8yr}
\end{figure}

We designate the BICEP3 95 GHz map
as ``L95'', in reference to the larger field-of-view/coverage area
(with respect to a \textit{Keck}-class receiver) and the 95 GHz
observing frequency. ``BK23 L95'' represents 5 additional seasons (2019--2023) of winter CMB
data as compared with the 3 seasons in BK18 L95, ultimately producing a map 
$\sqrt{\frac{3}{8.8}}$ times less noisy (8.8 rather than 8 owing to the suboptimal
performance of BICEP3 in 2016). The BK23 L95 (BICEP3 95 GHz 8-year) map is shown in figure~\ref{fig:b3_8yr}.
\begin{figure}[t]
  \centering
  \includegraphics[width=1\textwidth]{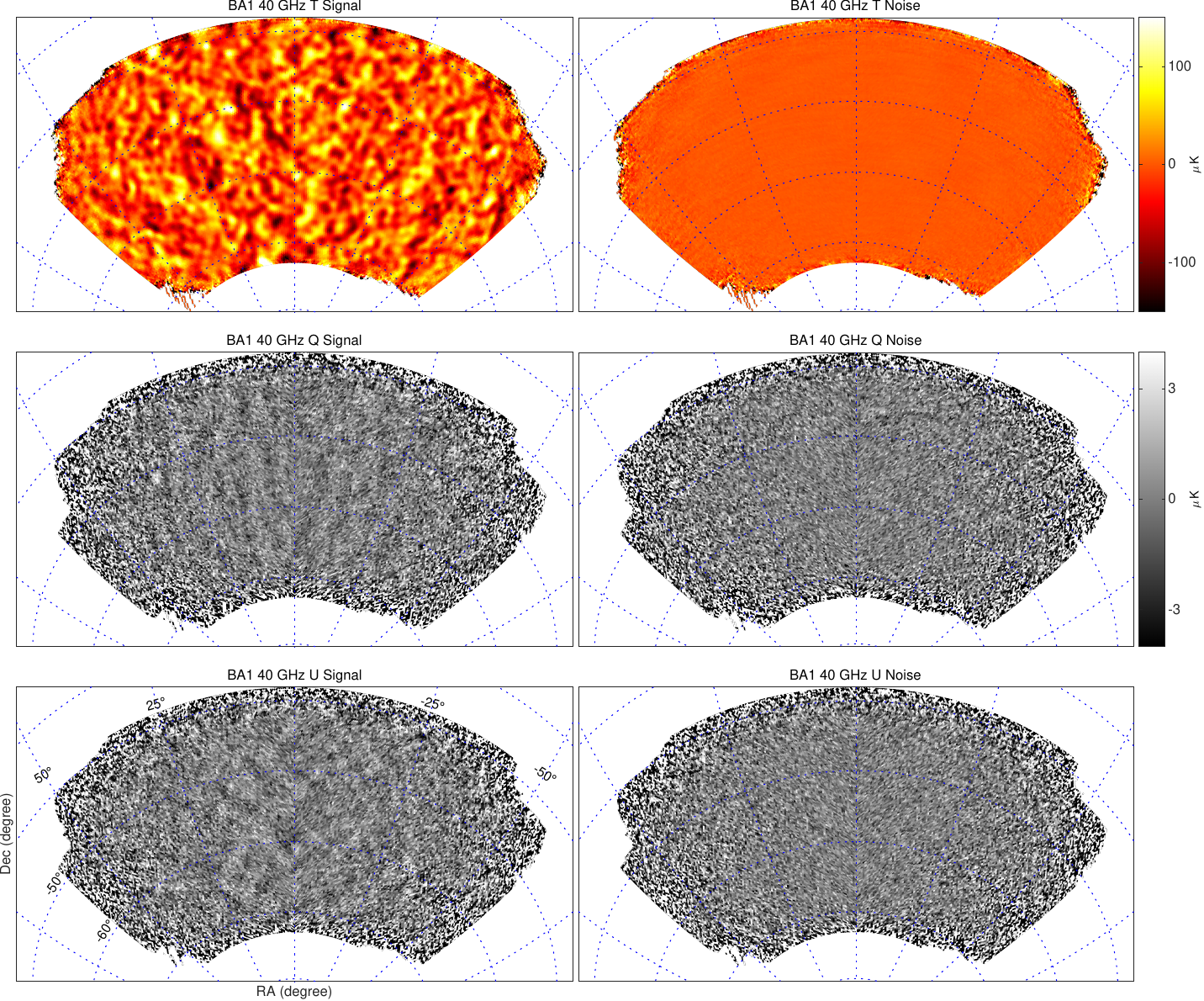}
  \caption{Preliminary BICEP Array 40 GHz 4-year (``BK23 L40'') temperature and
    polarization maps. From the top, the rows show temperature, Stokes
  $Q$, and Stokes $U$ maps. The left-hand column shows the signal
  maps, while the right-hand column shows a single realization of
  noise. The ``plus'' pattern in $Q$ and ``cross''
  pattern in $U$ is easily visible by-eye in the signal maps, and is
  characteristic of an E-mode-dominated sky.}
  \label{fig:ba1_40}
\end{figure}
We caution that this map should still be considered preliminary,
though note that we anticipate ultimate sensitivity levels better than
that which would be inferred solely from the fraction of additional data,
owing to improvements to data processing and analysis described in
section~\ref{subsec:reanalysis}.

\subsection{BICEP Array}
\label{subsec:ba}

The 30/40 GHz ``BA1'' receiver was deployed during the
2019--2020 Austral summer as well as the new BICEP Array telescope
mount. At the same time, three \textit{Keck} receivers, two observing
at 220 GHz and one at 270 GHz, were moved from the old mount, having
been installed in adapters to mimic the mass and
center-of-gravity of BICEP Array-class receivers.
BA1's focal plane is comprised of 12 detector module units, with
modules operating at either 30 or 40 GHz arranged in a
checkerboard pattern. The exact proportion and arrangement of modules
has changed over the first few years of operation. The 40 GHz modules
consist of a 5$\times$5 grid of detector pairs, while
30 GHz modules have a 4$\times$4 arrangement owing to the larger antenna
size. BA1's design and initial performance are described in further
detail in Schillaci \textit{et al.} (2020)~\cite{aleba1}. BA1
represents the first extension of \bk's frequency coverage below 95
GHz; the low-frequency channels are intended to aid in characterizing
the emission of galactic synchrotron which increases towards lower frequencies.
\begin{figure}[t]
  \centering
  \includegraphics[width=1\textwidth]{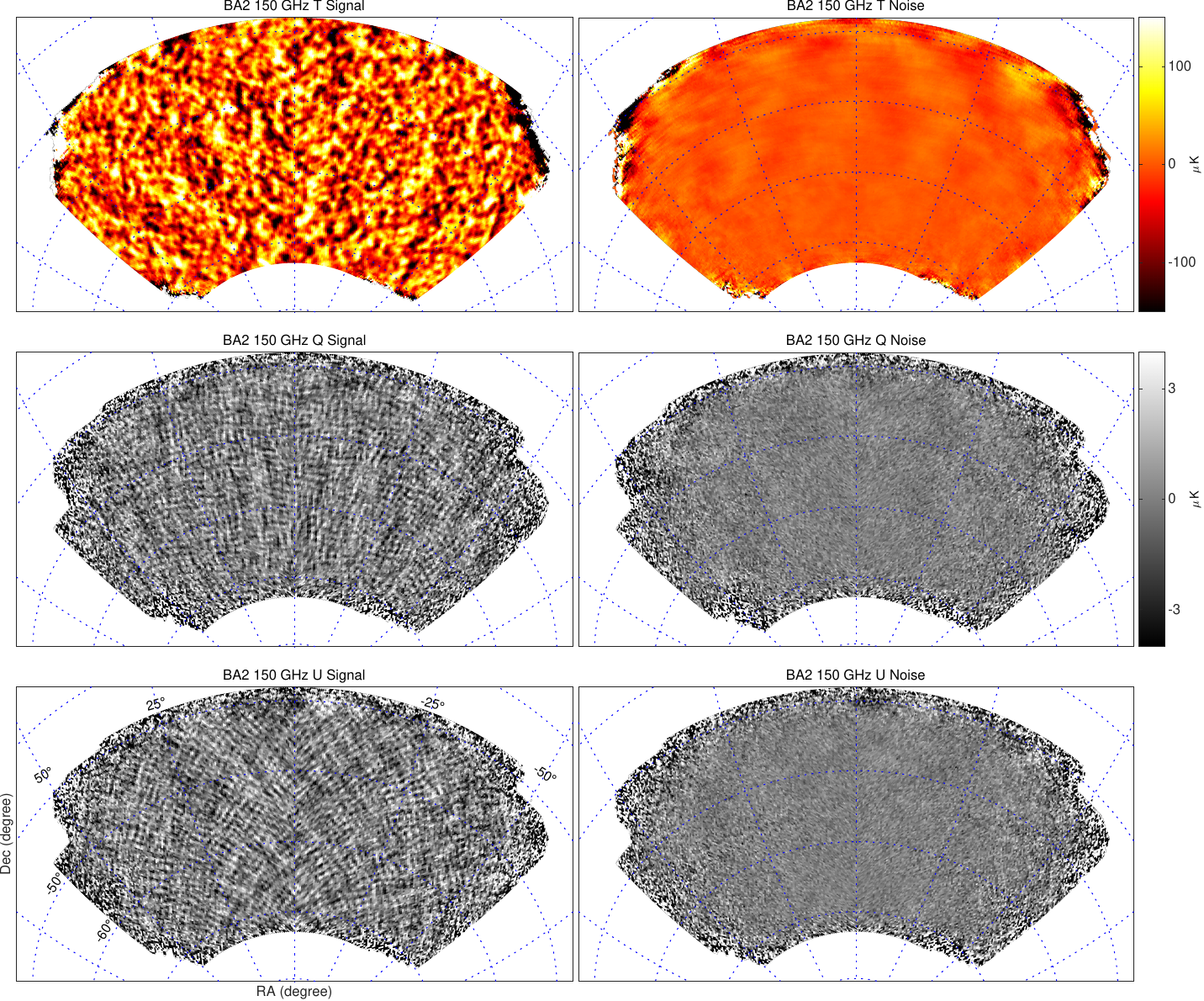}
  \caption{Preliminary BICEP Array 150 GHz 1-year (``BK23 L150'') temperature and
    polarization maps.}
  \label{fig:ba2_1yr}
\end{figure}

The restriction of Antarctic operations during the 2020--2021 Austral
summer caused by the COVID-19 pandemic resulted in BICEP Array
operating in the same configuration in 2021 as it did in 2020. During
the 2021--2022 Austral summer, a small team was able to deploy to
install new detector tiles in BA1, and to install new
optical filtering elements to eliminate unexpected out-of-band
coupling~\cite{ahmedspie}. The detector module upgrades represented a
shift towards a higher proportion of 30 GHz detector modules, intended
to improve sensitivity to the synchrotron foreground, and the
installation of some experimental dichroic
modules~\cite{corwindualband}.

Poor performance of the tiles installed in the 2021--2022
summer led to further focal plane upgrades to BA1 during the
2022--2023 and 2023--2024 summers. The 2022 30 GHz modules
were replaced with better-performing ones, and the 2022
dichroic tiles were replaced with with known-well-functioning 40 GHz modules. As of the
2024 observing season, BA1 is comprised of 7 modules (175 pairs/350
TES detectors) at 40 GHz, 4 modules (64 pairs/128 TES detectors) at
30 GHz, and a single remaining dichroic module (16 each of 30 GHz and 40
GHz pairs). The BK23 L40 maps are shown in figure~\ref{fig:ba1_40}.

During the 2022--2023 Austral summer season, one of the 220 GHz
\textit{Keck}-class receivers was decomissioned and replaced by the
second BICEP Array-class receiver, ``BA2'', observing at 150 GHz. For
this initial observing season, only 5 of the 12 detector modules were
installed due to fabrication and testing throughput
limitations. These modules were arranged on the focal plane such that
the overall map coverage was similar to BICEP3/BA1, while maintaining
good redundancy of coverage over differing boresight angles. BA2 also
employed an ultra-thin HMPE vacuum window as was installed in BICEP3
for the 2023 season, which should be even more beneficial at the
higher observing frequency.

BA2 detector tiles consist of 18x18 grids of detector pairs, \textit{i.e.}
324 pairs (648 individual TES detectors) per wafer -- 25\% more than
in an entire \textit{Keck} 150 GHz receiver. BA2 will ultimately
observe with 12 of these tiles for a total of 3888 pairs (7776 TES
detectors). The design and development of the 150 GHz modules for BA2
is described in Schillaci \textit{et al.} (2023)~\cite{aleba2}.

The five tiles installed in BA2 in 2023 have a total of 1620
optically coupled detector pairs, already a 25\% increase in detector
count over BICEP3. In figure~\ref{fig:ba2_1yr}, we show the
preliminary BA2 first-year (``BK23 L150'') temperature and
polarization maps. Five additional modules were installed in BA2 during
the 2023--2024 Austral summer, bringing the current total to 10 modules/3240
detector pairs (6480 TES detectors).

In addition to the introduction of BA1 and BA2, BK23 will contain 10
additional \textit{Keck}~receiver-years of 220 GHz data taken in 2019
and with the \textit{Keck}~receivers installed in BICEP Array
(representing a $\sim50\%$ increase over BK18), and the
first data (6 \textit{Keck}~receiver-years) at 270 GHz. This
high-frequency data will aid significantly in the discrimination of
the galactic dust foreground emission. However, owing to the smaller
instantaneous field of view, the maps produced by these
\textit{Keck}-class receivers will not provide information on
foregrounds for the additional sky area covered by BICEP3/BA-class
receivers. This capability, as well as a significant increase in detector
count, will be provided by the deployment of the 220/270 GHz
``BA4'' receiver, described further in subsection~\ref{subsec:ba4}.

\section{Future Outlook}
\label{sec:future}

\subsection{BA4}
\label{subsec:ba4}

The third BICEP Array-class receiver to be deployed is designated
``BA4''. BA4 will observe at 220 and 270 GHz, again employing
monochromatic (220 or 270 GHz) detector modules, 6 at each band, arranged
in a checkerboard pattern. BA4 will be time division-multiplexed with
the same architecture as is used in BA2, and will therefore have an
identical density and number of detectors (324 detector pairs
per module $\times$ 12 modules). BA4, through its larger field-of-view
and vastly increased detector count, will significantly improve
measurements of the galactic dust foreground, and its increased
resolution and sensitivity may even
enable more sophisticated studies of the distribution of polarized
dust in the galactic plane. BA4 is currently being integrated and undergoing in-lab calibration
measurements at Stanford University, and is planned to deploy to the
South Pole this coming 2024--2025 Austral summer.

%\begin{wrapfigure}{r}{0.25\textwidth}
%  \centering
%  \includegraphics[width=0.25\textwidth]{ba4}
%  \caption{The BA4 receiver in-operation at Stanford University,
%    conducting Fourier transform spectrometer measurements of detector
%  passbands.}
%  \label{fig:ba4}
%\end{wrapfigure}

\subsection{Reanalysis}
\label{subsec:reanalysis}

A key feature of the upcoming BK23 map set will be the
from-scratch reanalysis of all archival data dating back to BICEP2's
2010 observing season. Previously, new seasons of data were analyzed
and validated independently, before being coadded to the previous
dataset's maps, thus locking-in choices, techniques, and
software initially chosen over a decade ago. A key philosophy behind
the implementation of this reanalysis project was optimizing
performance, scalability, and compatibility, to enable easier
collaboration with external groups (such as adopting the widely-used
HEALPix~\cite{healpix} pixelization) and more rapid iteration of
analysis choices.
\begin{figure}[t]
  \centering
  \includegraphics[width=\textwidth]{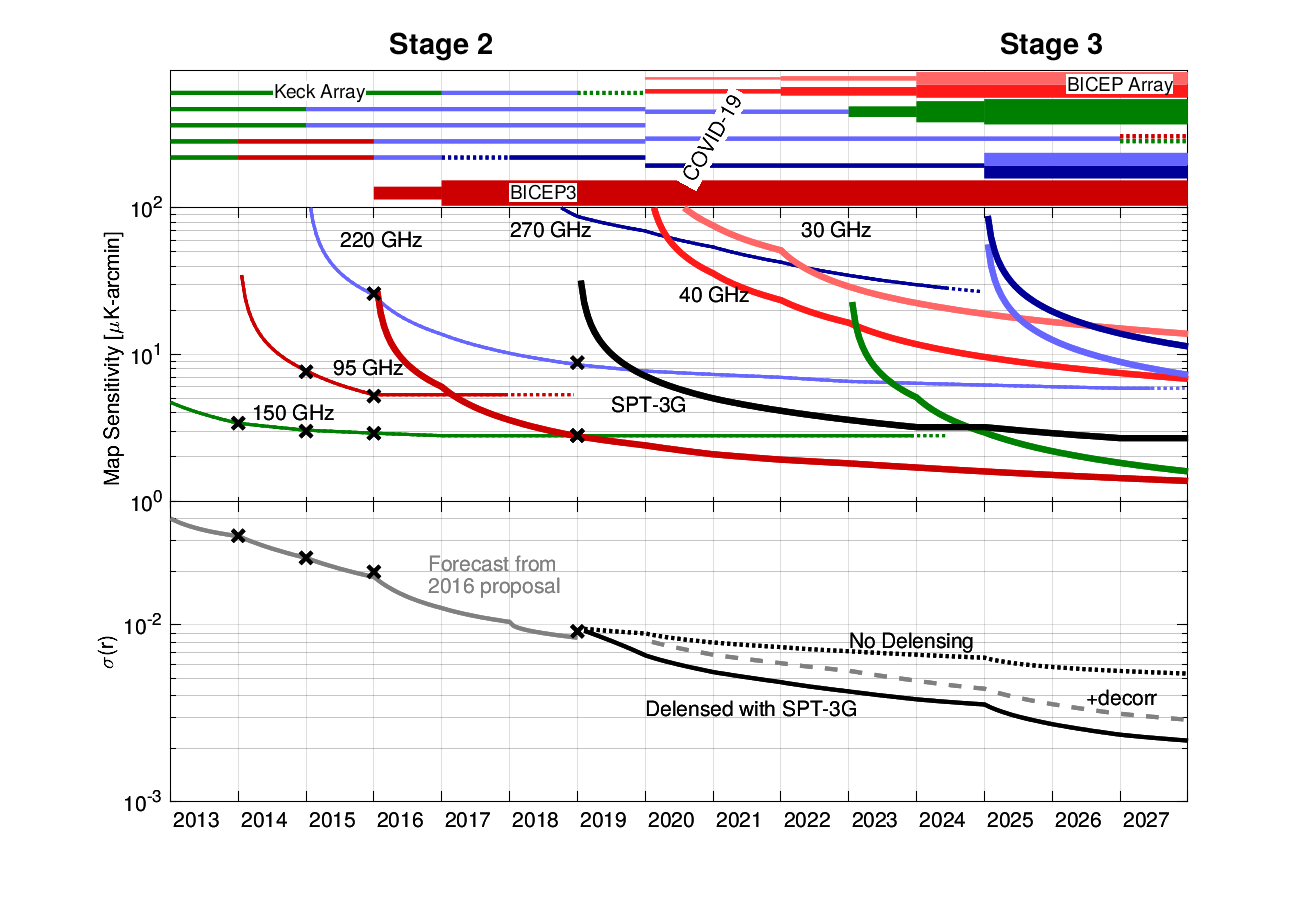}
  \caption{Achieved and projected sensitivity of the \bk\ experiment
    and its constituent observing bands. The top subfigure shows a
    schematic of the observing configuration during a given season,
    with line color being a proxy for observing frequency and
    width for detector count. The middle subfigure shows the
    progression of raw sensitivity in each BK observing band, and that of
    SPT-3G (solid black) used for delensing. The bottom subfigure shows
    the progression of the primary figure-of-merit for the experiment,
    $\sigma(r)$, as a function of time; the solid line denotes the
    sensitivity when delensing in conjunction with SPT-3G (removing
    about 70\% of the lensing signal at full sensitivity), while
    the dotted represents the progression of
    sensitivity with no delensing --- note the much earlier level-off
    of $\sigma(r)$ if delensing is not performed. The gray dashed line
    is the result of freeing a parameter used to model
    first-order decorrelation of foregrounds as a function of
    frequency. Black ``$\times$'' denote published values.}
  \label{fig:projplot}
\end{figure}

Low-level changes implemented for the reanalysis include optimizations
to: readout transfer function deconvolution, the removal of timestream glitches, timestream
filtering, timestream weighting, and data quality cuts, among others. Timestream
polynomial filtering has been moved to a Legendre polynomial basis to
better orthogonalize the removed modes, scan-synchronous subtraction
has been moved to a longer timescale, and timestream weighting has
been optimized to prioritize the low-frequency regime of interest for
constraining $r$.

Alongside low-level processing changes, a new automated data reduction
pipeline and suite of automatically generated diagnostics has been
constructed. This reduction pipeline emphasizes
transparency, robustness, and ease-of-use, while the diagnostics
represent significant improvements in information density and
accessibility. The ability to accurately assess increasingly large
volumes of incoming data is critically important to overall observing
efficiency, and will only become more so as new experiments with even
larger detector counts come online.

One more notable area of improvement in the reanalysis is the
overhauling of internal consistency testing and map validation. This
includes varying and in many cases extending the amount of data used
in various data splits used to probe for systematic effects, and
constructing more robust statistical tools to assess the outcomes of
those tests.

\subsection{Delensing}
\label{subsec:delensing}

Figure~\ref{fig:projplot} shows the achieved and projected reach
of the \bk\ experiment over time. A noteworthy feature of the
$\sigma(r)$ plot is that sensitivity rapidly levels-off despite
increasing per-band map depths if there is no delensing. In BICEP's
relatively small patch, the number of lensing modes is limited, and
thus the effective sample variance on lensing is large; indeed, as
mentioned it is already the dominant contribution to BK18's
$\sigma(r)$, rather than uncertainty on foreground emission. With
detailed high-resolution polarization maps and knowledge of the
integrated lensing potential provided via BICEP's collaboration with
the South Pole Telescope (SPT) --- constituting the South Pole Observatory
(SPO) --- the specific lensing modes can be subtracted, enabling much
deeper $r$ constraints even on a small patch. This delensing approach
has already been demonstrated successfully on older BK, SPT, and
\textit{Planck} data~\cite{kimmydelensing}, and new
techniques~\cite{juliandelensing} promise even further
improvements. BICEP's current results and the sensitivity levels of
newly collected data all affirm the necessity of delensing to achieve competitive
limits on $r$ going forward.

\section*{Acknowledgments}

\bk~has been made possible through U.S. National Science Foundation, most recently including 2220444-2220448, 2216223, 1836010, and 1726917.  We also thank our heroic winter-over operators: Manwei Chan, Karsten Look, Calvin Tsai, Paula Crock, Ta- Lee Shue, Grantland Hall, Hans Boenish, Sam Harrison, Anthony DeCicco, Thomas Leps, Brandon Amat, and Nathan Precup, Steffen Richter, and Robert Schwarz.

\end{document}